\documentstyle[11pt,aaspp4,flushrt]{article}  

\def\msun{{\rm M_{\odot}}}

\def\sax{SAX J1808.4-3658 }
\begin{document}

\title{A FIRM UPPER LIMIT TO THE RADIUS OF THE NEUTRON STAR IN \sax}

\author{L.~Burderi\altaffilmark{1} and A.~R.~King\altaffilmark{2}}
\authoremail{burderi@gifco.fisica.unipa.it, ark@star.le.ac.uk}

\altaffiltext{1} {Dipartimento di Fisica ed Astronomia, Universit\`a di 
Palermo, via Archirafi 36, 90123 Palermo, Italy}

\altaffiltext{2} {Astronomy Group, University of Leicester, 
Leicester LE1 7RH, U.K.}

\begin{abstract}
We show that observations of X--ray pulsing from \sax place a firm
upper limit of $13.8 m^{1/3}~{\rm km}$ on the radius of the neutron
star, where $m$ is its mass in solar units. The limit is independent
of distance or assumptions about the magnetospheric geometry, and
could be significantly tightened by observations of the pulsations in
the near future. We discuss the implications for the equation of state
and the possible neutron star mass.

\end{abstract}


\section{INTRODUCTION}

The discovery (Wijnands and van der Klis, 1998) of an accreting neutron
star 
with a 2.49 ms spin period in the soft X--ray transient \sax 
(orbital period = 2.01 hr, Chakrabarty and Morgan, 1998)
has important consequences for many aspects of our understanding of
low--mass X--ray binaries and their evolution. In this Letter we focus
on just one of these, namely the rather tight limit one can place on
the neutron star radius. This in turn has significant implications for
the equation of state and the possible mass of this star.

\section{THE RADIUS LIMIT}

Our radius limit comes from considering the fact that pulsed X--ray
emission is detected at a range of X--ray count rates. 
We make a number of assumptions, which are all commonly adopted by
other authors; later in the paper we shall consider the
consequences of relaxing them. To specify them we define the 
neutron star radius $R_*$, 
the corotation radius
\begin{equation}
R_{\rm co} = 1.5\times 10^6m^{1/3}P_{-3}^{2/3}~{\rm cm}
\label{co}
\end{equation}
and the magnetospheric radius
\begin{equation}
R_{\rm M} = 1.9 \times 10^6 \phi \mu_{26}^{4/7} m^{-1/7}
\dot{M}_{17}^{-2/7}~{\rm cm},
\label{mag}
\end{equation}
({\it e.g.} Burderi {\it et al.}, 1998).
Here $m$ is the neutron star mass in solar masses, $P_3$ is the
neutron star spin period in milliseconds, $\phi\sim 1$ is the ratio 
between the magnetospheric radius and the Alfv\'en radius ({\it e.g.}
Burderi 
{\it et al.}, 1998), $\mu_{26}$ is the neutron star
magnetic moment in units of $10^{26}$ Gauss cm$^{3}$, $R_6$ is the neutron
star radius in units of $10^{6}$ cm, $\dot{M}_{17}$ is the accretion rate
in
units of $10^{17}$ g/s.
 
We derive a limit using the assumptions A1
-- A4 defined below. These are that X--ray pulsing requires 

A1: 
\begin{equation}
R_* < R_{\rm mag}
\label{A1}
\end{equation}

and 

A2:
\begin{equation}
R_{\rm mag} < R_{\rm co},
\label{A2}
\end{equation}
and in particular that there is no `intrinsic' pulse mechanism (as e.g. in
rotationally powered radio pulsars).

We assume also that

A3: In eqn (\ref{mag}), $\phi$ is independent of $\dot M$,

and

A4: the accretion rate 
$\dot M$ is strictly proportional to the X--ray count
rate $F$ (3 -- 25 keV band) in the Proportional Counter Array (PCA) 
instrument on RXTE.

Using A4 and A3 we can simplify eqn (\ref{mag}) to
\begin{equation}
R_{\rm mag} = AF^{-2/7},
\label{F}
\end{equation}
where $A$ is a constant for a given system.

X--ray pulsing was detected from \sax at count rates corresponding to
$F_1 = 238 \times 10^{-11}$ erg/cm$^{2}$/s and $F_2 = 21 \times 10^{-11}$ 
erg/cm$^{2}$/s (Gilfanov {\it et al.} 1998). Applying A1, A2 and (\ref{F}) 
to these observations gives
\begin{equation}
R_* < AF_1^{-2/7}
\end{equation}
and
\begin{equation}
AF_2^{-2/7} < R_{\rm co}
\end{equation}

But $AF_2^{-2/7} = 2.00 \times AF_1^{-2/7}$, so by combining the two we get the
extremely simple result
\begin{equation}
R_* < 0.5 R_{\rm co} = 13.8 \times m^{1/3}~{\rm km}
\label{lim}
\end{equation}
Usually our lack of knowledge of the uncertain quantity $\phi$ and to
some extent $\dot M$ frustrates
attempts to get precise quantitative information from assumptions like
A1, A2. However here all of these uncertainties are removed because of
the scaling (\ref{F}): given the assumptions A1 -- A4, the upper limit
(\ref{lim}) depends {\it only} on the neutron star mass.

\section{IMPLICATIONS}

Neutron--star equations of state generally yield a theoretical
mass--radius relation of the form
\begin{equation}
R_* = Cm^{-1/3}.
\end{equation}
When combined with such a relation the limit (\ref{lim}) defines a
minimum mass for the neutron star in \sax, while the characteristic
lengthscale $L$ measured from the Type I X--ray bursts (in't Zand 
{\it et al.}, 1998)
defines maximum mass via the condition $L < R_*$ (i.e. the burst site
cannot exceed the entire neutron--star surface).
%
%
This last condition is derived assuming that the peak luminosity
during the X--ray bursts is close to the Eddington luminosity:
\begin{equation}
L = 31.3 \times (kT/1{\rm keV})^{-2} m^{1/2}~{\rm km} < R_*.
\label{burst}
\end{equation}
In applying equation (\ref{burst}) two points have to be considered. 
Firstly,
the blackbody temperature ($kT = 2.23$ keV) was determined by averaging the 
values obtained from the first 7 sec of the burst reported in 't Zand 
{\it et al.} (1998) together with 
the first 7 sec of the second burst. During
the initial phase of the bursts photospheric expansion is likely, so
the photospheric radius $L$ could exceed the radius of the neutron star.
Secondly (\ref{burst}) 
is derived under the assumption of pure blackbody emission, while
more accurate determinations of the spectrum of the emitting photosphere
show that several corrections apply to the spectral shape (see {\it e.g.}
Titarchuk, 1994), resulting into a underestimate of the radius of the
photosphere. 
%
%
Figure 1 shows a mass--radius plane with these constraints marked for
various equations of state. For any reasonable equation of state 
the neutron star radius must be
smaller than 15~km. As can be seen, very stiff equations of
state (EOS) (such as the Mean Field EOS labeled as L see {\it e.g.} 
Cook, Shapiro and Teukolsky, 1994)
are marginally allowed if the $m$ takes the favoured
value of 1.4, and softer equations of state (such as the Reid soft core
EOS,
labeled A or the FPS EOS see {\it e.g.} 
Cook, Shapiro and Teukolsky, 1994) can currently accomodate
even smaller masses. However the upper radius limit (\ref{lim}) is at
present only derived for X--ray count rates down to $F_2 = 21 \times 
10^{-11}$ erg/cm$^{2}$/s. The system is
detected at much lower rates, and it will be possible to check for
pulsation and thus reduce the
allowed area in the $R_* - m$ plane quite drastically in the very near
future. The Figure shows the resulting constraints if pulsing is still
detectable at count rates $9 \times 10^{-11}$ 
erg/cm$^{2}$/s and $3 \times 10^{-11}$ 
erg/cm$^{2}$/s. The detection of
pulsing at a count rate $3 \times 10^{-11}$ 
erg/cm$^{2}$/s would force $M$ to be significantly larger
than 1.4 for even the softer equations of state. This would be quite
reasonable, given that the formation mass for the secondary star in 
neutron--star LMXBs is probably $\ga 1{\rm M}_{\odot}$ (cf King \&
Kolb, 1997, Kalogera, Kolb \& King, 1998), and that significant mass
must have been transferred to the neutron star during its evolution.

\section{DISCUSSION}

Here we consider the effect of dropping each of the assumptions A1 --
A4. 

A1 is normally regarded as the minimum condition for the magnetic
field to be dynamically important. Pulsing while this condition was
violated would suggest that the neutron star had some intrinsic pulse
mechanism. This is very unlikely to be the usual radio--pulsar
process, as this is known to be effectively extinguished if there is
significant matter inside the light cylinder, as would be true
in this case. Moreover the idea of such an intrinsic component
taking over from the normal accretion power--pulsing at low count
rates is belied by the almost constancy of the pulse fraction (Gilfanov
{\it et al.}, 1998)

Dropping A2 would imply that pulsing was possible in the `propellor'
regime, when matter is thought to be centrifugally expelled rather
than accreted. Again it is hard to reconcile this idea 
with the constancy of the observed X--ray pulse fraction.

Assumption A3 appears to be perhaps surprisingly accurate, considering that
it is only a dimensional estimate that is given by comparing magnetic and
material stresses.
%
%
Campbell (1997) derives a general prescription for the disruption 
radius $R_{\rm mag, disc}$ of a thin disc of central density $\rho$ by a
dipole aligned with the spin axis
\begin{equation}
R_{\rm mag, disc} \propto \rho^{1/9}\mu^{2/3}\dot M^{-4/9}.
\label{camp}
\end{equation}
The weak $\rho$--dependence in this relation means that 
if disruption occurs in a region of
the disc where gas pressure is dominant and thus $\rho \propto
\dot M^{1/2}R^{-3/2}$ (cf Frank et al., 1992) one gets 
$R_{\rm mag, disc} \propto \dot M^{-1/3}\mu^{4/7}$, 
i.e. quite close to the dimensional estimate (\ref{mag}) with
$\phi \sim$ constant. If instead radiation pressure is important 
near the disruption radius one
has $ \rho \sim \dot M^{-2}R^{3/2}$, which would give the stronger
dependence 
\begin{equation}
R_{\rm mag, disc} \sim \dot M^{-4/5}\mu^{4/5}.
\label{rad}
\end{equation}
However  
the rather low luminosity ($\la 10^{36}$~erg~s$^{-1}$) of \sax implies
accretion rates $\dot M \la 10^{16}$~g~s$^{-1}$, so that radiation
pressure only becomes comparable with gas pressure 
for disc radii $R \la 20$~km (Frank et al.,
1992), as compared with the likely corotation radius $R_{\rm co}
\simeq 30$~km. The disruption therefore probably does not occur in a
region dominated by radiation pressure. However for comparison we plot the
curve for $F_2 = 21 \times 10^{-11}$ erg/cm$^{2}$/s
with the dependence (\ref{rad}) in Figure 1. As is possible to observe,
this curve is well below the lower limit imposed by (\ref{burst}), supporting
our hypothesys that the disc is truncated in a zone where radiation 
pressure is negligible.

%
%

The X--ray spectrum of \sax is remarkably constant (Gilfanov
{\it et al.}, 1998), suggesting
that A4 is unlikely to be upset by a changing bolometric correction. A
more serious possibility is that some of the accretion energy might be
stored in other forms, rather than being released promptly as
radiation. For example Priedhorsky (1986) suggested that the neutron
star spin might provide such a reservoir in the QPO sources. To weaken
the limit (\ref{lim}) significantly would require essentially that the
drop in X--ray count rate masked an almost constant accretion
rate. Again the constancy of both the X--ray spectrum and the pulse
fraction tend to go against this idea. 

We conclude that the limit {\ref{lim}) is robust. The work of this
paper also shows that the magnetospheric radius in \sax lies between 1 and
at
most a few times $R_*$. While it is possible that this is merely a
coincidence, it may instead suggest that whatever process induces
field decay tends to stop once such conditions are achieved. This in
turn supports the idea that the decay may be induced by accretion.

After submitting our paper for publication we became aware of an
independent paper by Dimitrios Psaltis and Deepto Chakrabarty which reaches 
similar conclusions.

\acknowledgements
This work was supported
by the UK Particle Physics and Astronomy Research Council through a Senior
Fellowship (ARK) and a Visiting Fellowship Grant to the Leicester 
Astronomy Group.

\clearpage
 
\section*{FIGURE CAPTIONS}
\bigskip

Constraints on the radius of the neutron star in \sax. This is bounded
above by the solid curve labelled $7.0\times 10^{35}$~erg~s$^{-1}$,
corresponding to the lowest X--ray luminosity at which pulsing is
currently detected (and derived adopting the assumption A3, see text), 
and bounded below by the size derived from the
observation of Type I X--ray bursts (lower solid line). Detections of
pulsing at $3.0\times 10^{35}$~erg~s$^{-1}$ or $1.0\times
10^{35}$~erg~s$^{-1}$ would reduce the upper limit to that given by the
appropriate dashed curve. 
For comparison, the upper limit derived adopting the disruption radius 
dependence (\ref{rad}) and the luminosity $7.0\times 10^{35}$~erg~s$^{-1}$
is shown by the curve labelled Radiation pressure limit.
The curves `L', `FPS' and `A' give the
mass--radius relations predicted by the equations of state discussed 
in the text.
For each such equation, the upper radius limit
defines a maximum radius and minimum mass for the neutron star. Thus
equation of state L currently requires $M \ga 1.34\msun$, $R \la 15$~km.
The latter is probably the maximum reasonable radius for the neutron star
in \sax.

\bigskip
\bigskip

\clearpage


\begin{thebibliography}{}
\bibitem{}
Burderi, L., T. Di Salvo, N. R. Robba, S. Del Sordo, A. Santangelo,
A. Segreto, 1998, ApJ, 498, 831
\bibitem{}
Campbell, C.G., 1997, Magnetohydrodynamics in Binary Stars (Dordrecht,
Kluwer Academic Publishers)
\bibitem{}
Chakrabarty, D., Morgan, E. H., 1998, Nature, submitted
\bibitem{}
Frank, J., King, A.R., Raine, D.J., 1992, Accretion Power in
Astrophysics, 2nd Ed. (Cambridge: Cambridge University Press)
\bibitem{}
Gilfanov, M., Revnivtsev, M., Sunyaev, R., Churazov, E., 1998, 
astro-ph/9805152 
\bibitem{}
Cook, G. B., Shapiro, S. L., Teukolsky, S. A., 1994, ApJ, 424, 823
\bibitem{}
King, A. R., Kolb, U., 1997, ApJ, 481, 918
\bibitem{}
Kalogera, V., Kolb, U., King, A. R., 1998, ApJ, in press
\bibitem{}
Priedhorsky, W., 1986, ApJ, 306, L91
\bibitem{}
Titarchuk, L., 1994, ApJ, 429, 340
\bibitem{}
Wijnands, R., van der Kils, M., 1998, Nature, submitted
\bibitem{}
in't Zand, J. J. M., {\it et al.}, 1998, A\&A, 331, L25
\end{thebibliography}
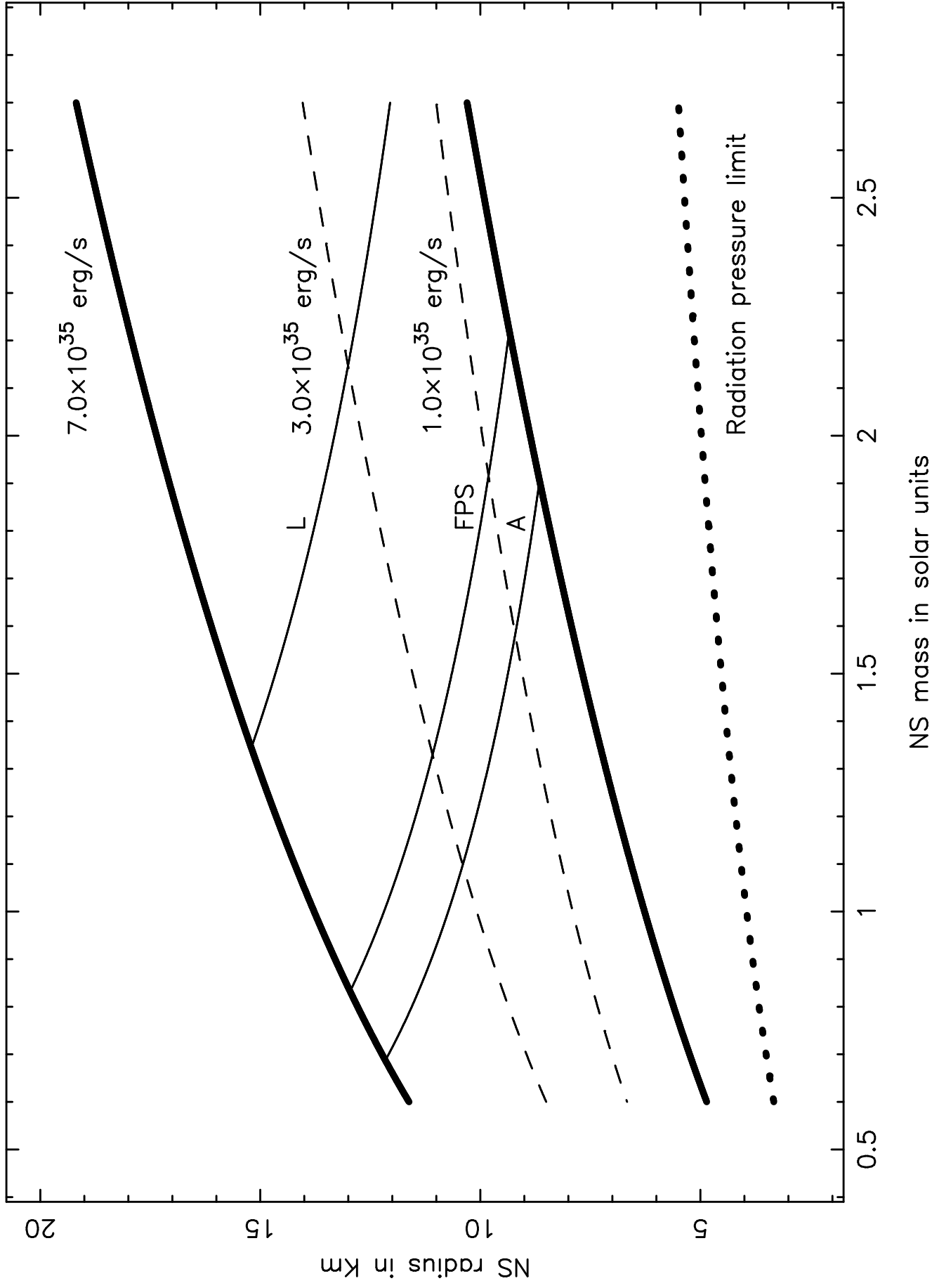
\end{document}